\documentclass[aps,prl,twocolumn,showpacs,superscriptaddress]{revtex4-1}
\usepackage{amsmath}
\usepackage{epsfig}
\usepackage{amssymb}
\usepackage{dcolumn}
\usepackage{graphicx}
\usepackage{dcolumn}

\usepackage{color}

\newcommand{\beq}{\begin{equation}}
\newcommand{\eeq}{\end{equation}}

\begin{document}
\date{\today}

\title{Nonlinear Eigenmodes of a Polariton Harmonic Oscillator}

\author{Florian Pinsker}
\email{florian.pinsker@gmail.com}

\affiliation{Department of Applied Mathematics and 
Theoretical Physics, University of Cambridge, United Kingdom.} 

\author{Tristram J. Alexander}
\email{t.alexander@unsw.edu.au}

\affiliation{School of Physical, Environmental and Mathematical Sciences, UNSW Canberra, Canberra ACT 2600, Australia}



\begin{abstract}
We investigate theoretically the quantum oscillator-like states recently observed experimentally in polariton condensates (Nat. Phys. {\bf 8}, 190 (2012)).  We consider a complex Gross-Pitaevskii type model which includes the effects of self-interactions, and creation and decay of exciton-polaritons.  We develop a perturbation theory for approximate solutions to this non-equilibrium condensate model and compare the results with numerically calculated solutions for both repulsive and attractive polariton-polariton interactions.  While the nonlinearity has a weak effect on the mode selection their density profiles are modified at moderate gain strengths and becomes more dominant when a very large gain of polaritons implies an extended cloud  with high condensate densities.  
Finally, we identify the relation of the observed patterns to the input pump configuration, and suggest this may serve as a generalized NOR gate in the tradition of optical computing.
\end{abstract}


\maketitle

\section{Introduction}

Bose-Einstein condensates (BECs) are prime examples of {\it nonlinear} quantum many-body systems, because of interactions between coherent quantum particles.  They are regarded as a macroscopic quantum phenomenon due to their very high coherence and long-range order of all particles constituting this state of matter, a quantum state that typically is formed at very low temperatures \cite{E, 1,2,3, Stri, Light}. Many setups of atomic BEC are for fixed particle numbers and can carry various topologically stable excitations ranging from solitons to giant vortices \cite{2,PRA,Corr, Corr2,dark}. More recently efforts have been made towards the realization of coherent {\it non-equilibrium} many-body systems that exchange particles and/or energy while at the same time obeying BEC properties e.g. in solid state systems \cite{Kasp, Baas}. One such system is the polariton quasiparticle Bose-Einstein condensate \cite{Kasp, Baas, 1,3,Light}, which can be effectively described by a non-conservative nonlinear Schr\"odinger-type equation \cite{Light, Pin9, Pin11} formally generalizing the so called Gross-Pitaevskii equation (GPE) - the model for the condensate wave function of dilute and weakly interacting atomic BEC \cite{Stri, Pin9,Pin11}.  

Bose-condensed polaritons have become feasible in semiconductor microcavities in the last decade \cite{Baas, Light, 1,Kasp}.  Polaritons in semiconductor microcavities are superpositions between excitons (electron-hole pairs) and cavity photons, and as such the combination of the components determines the particle statistics \cite{Nat1,1}.  These quasiparticles in a dilute regime show distinctive properties such as Bose-Einstein condensation,  superfluidity, a  finite nonzero speed of sound due to the nonlinearity and the emergence of elementary topologically stable excitations such as solitons or quantum vortices \cite{Pin6,Pin9,Pin10,1,Light}. In their very nature polariton BECs are non-equilibrium systems - polaritons can be created via a local light field and decay after $1$-$100$ ps mainly due to leaking of the cavity photons \cite{Kasp,Light, snoke}.  The decay rate significantly depends on the quality of confinement of the light field \cite{snoke} within the microcavity and in this work it is assumed to be constant over space.  From a technological and experimental point of view polariton BECs have several practical advantages over atomic BECs.  For instance, BECs of dilute atoms must be realized at temperatures within the nano-Kelvin range \cite{first,second, third}, whereas polariton condensates can be observed even at room temperature in polymers \cite{Plum} and in organic materials \cite{organic}. This is a consequence of the fact that polaritons have a very light mass due to their inherent coupling to photons \cite{Light}, and so have a much higher critical Bose condensation temperature.  Restrictions on the polariton density however have to be taken into account to ensure their bosonic character is maintained \cite{2}.  

In many settings a polariton condensate is an effectively $2$D coherent many-body system that is tightly confined with regards to the optical axis of the semiconductor microcavity  \cite{Baas,Kasp,Pin8,1}. Depending on additional constraints the condensate may become essentially $1$D \cite{Pin6,1}.  The elementary excitations within a $1$D system, with repulsive polariton-polariton interactions, are dark and grey solitons \cite{Pin6,Light}. Grey solitons are stable excitations which move over long distances without changing their form and in principle with elastic scattering properties, but due to the non-equilibrium character obey a finite lifetime \cite{Malu}. Dark solitons are the stationary special case of the grey soliton when the density depletes to zero at its minimum and with a $\pi$ phase shift at the nodal point \cite{Stri}. 

In this work we are particularly interested in stationary and stable patterns within a $1$D polariton condensate, that once excited are similar to dark soliton \cite{Pin9} arrays \cite{Pin6} with respect to singularities at localized points  - the oscillator modes of a harmonically trapped quantum state. Experimental observations of oscillator modes in polariton condensates have been made in \cite{Tosi,Hamid}, but a thorough theoretical description is lacking. The experimentally observed non-equilibrium  quantum oscillator states were formed between two spatially separated pump spots of incoherently pumped polaritons, corresponding to two localized laser beams shining on the semiconductor microcavity and positioned within the $2$D plane. Between the pump spots stationary spatial patterns were formed similar to those observed for a quantum harmonic oscillator, hence suggesting an effective $1$D problem. 

The spatial patterns can be understood in a dynamical picture as follows. The polaritons generated at the spots move from the source peaks in all directions, because interactions of pump and reservoir with polaritons are repulsive \cite{Pin8}. So in between the pump spots the bulk of polaritons are present and thus the condensate and its patterns form predominantly there. Depending on the pump strength, and the repulsiveness from the pump/reservoir, different modes can be observed; starting from the ground state of approximately gaussian form (when nonlinearity plays a minor role), excited states become feasible. Experimentally only the case for repulsive self-interactions has been considered so far. In this work we predict - based on the recent experimental effort \cite{nature} - the states for attractive self-interactions as well and their dependence on the gain. To elucidate the role of nonlinear self-interactions for the pattern formation in quasi-harmonically trapped polariton condensates we vary between attractive and repulsive interactions and the nonlinear character of polariton BEC will become dominant in a Thomas-Fermi (TF) type regime, i.e. when condensate densities are very high and pumping is strong.

Our paper is organized as follows. First we will introduce the general model describing the polariton condensate in $2$D and a simplified model for the effective $1$D investigations. Then approximate analytical expressions for the condensate wave functions are presented and compared with numerical simulations. Subsequently the dependence of the eigenvalues on pumping strength and the nonlinearity is presented.  Finally we introduce a generalized optical NOR device and summarize our findings.

\section{Polariton condensate model} 

We describe the order parameter $\psi$ of the polariton condensate by the state equation in the spin coherent case \cite{W,Pin11}, which has proven to be an accurate description of recent experiments \cite{Pin8, Pin9},
\begin{multline}\label{nat}
 i \partial_t  \psi = \bigg( - \frac{\hbar \nabla^2}{2 m^*} + g_R N  + \\ +  g_P P + \alpha_1 |\psi|^2   +  \frac{i }{2} \left(\gamma_R  N - \gamma_C  \right) \bigg) \psi.
\end{multline}
where $m^*$ is the effective mass of the condensed polaritons and $P$ describes the geometry and magnitude of the pump spots, which are induced by linearly polarized laser beams \cite{Hamid,Tosi}. The polariton reservoir is approximately given by  \cite{Pin8,Pin9}
\beq
N \simeq \frac{P}{\gamma_R} \left(1- \frac{\beta}{\gamma_R}   |\psi|^2 \right).
\eeq  
In Eq. \ref{nat} $\gamma_C$ denotes the globally constant decay rate of condensed polaritons, $\gamma_R$ the relaxation of the reservoir, $\beta$ is the scattering rate of polaritons into the condensate, $\alpha_1$ denotes the polariton-polariton interactions within the condensate, $g_P$ is the repulsive interaction of the condensate with the pump and $g_R$ the repulsive interactions of the condensate with the reservoir. Parameter values are chosen in accordance to most recent experiments \cite{Pin6,Pin9}, i.e. effective mass $m^*=5\times10^{-5}m_0$ where $m_0$ is the mass of the free electron, $\alpha_1 = 0.001  \cdot ps^{-1} \cdot  \mu m^{2}$, $
g_R   = 0.022 \cdot ps^{-1}  \cdot \mu m^{2}$, $g_P   = 0.07 \cdot ps^{-1}  \cdot \mu m^{2}$, $
\beta  = 0.05  \cdot ps^{-1}  \cdot  \mu m^{2}$, $
\gamma_R  = 10   \cdot  ps^{-1} $, $R_R= 0.06  \cdot ps^{-1}  \cdot \mu m^{2}$, $\gamma_C  = 0.556 \cdot ps^{-1}$. The pump geometry is assumed to be
\begin{multline}\label{spots}
P (x,y) = A \bigg( \exp \left( - 0.073 \cdot( (x+x_1)^2 +y^2) \right) + \\ +  \exp \left( - 0.073 \cdot( (x-x_2)^2 +y^2) \right) \bigg)
\end{multline}
with $A$ being the amplitude and the position of the two gaussian pump spots is at $x_2$ and $x_1$ respectively, which models the gain due to local illumination of the semiconductor microcavity. If we neglect spatial extension in $y$ we simply write $P(x)$.  We note that the exact mathematical form does not significantly change the physics under consideration and is an approximation of the actual spot form used in recent experiments \cite{Pin8,Pin9}.

Taking this into account, and as outlined in the next section in more detail, we can effectively reduce \eqref{nat} to a simplified model for the order parameter $\psi$ of the polariton condensate given by \cite{jona}
\begin{multline}\label{n1}
i \hbar \partial_t \psi = \bigg(- \frac{\hbar^2 \nabla^2}{2m^*}  + V(r)+U|\psi|^2 + i (\gamma_{\rm eff} - \Gamma |\psi|^2) \bigg)\psi.
\end{multline}
Here $U$ denotes the self-interaction strength, $\gamma_{\rm eff}$ the linear gain  and $\Gamma$ the nonlinear loss. The repulsive effect on the polaritons of the pump $P$ and the reservoir $N$, defined in $\eqref{nat}$, is assumed to be approximated by $V$.  For our analysis we are interested in stationary states and so set $i \hbar \partial_t \psi = \mu \psi$. As a simplification, we neglect the transverse spatial dimension, and so write $\psi(x,y) = \psi(x)$, and rescale $\eqref{n1}$ as in \cite{jona}, i.e. $V(x) = (\hbar \omega/2)(x^2/l^2)$ (however without effective transverse confinement), where $\omega$ is the oscillator frequency and $l = \sqrt{\hbar/(m^* \omega)}$ the corresponding length. Expressing lengths in units of $l$ and energy in units of $\hbar \omega$, and by substituting $\psi \to \sqrt{\hbar \omega/(2U)} \psi$, the result is
\begin{equation}\label{q}
\bigg(\partial_{xx} - x^2 + (\sigma_1+i \sigma_2) |\psi|^2 -  i \mu_2 \bigg) \psi =   -\mu_1 \psi
\end{equation}
with $\mu_1 =   \left( \frac{2 \mu}{\hbar \omega} \right)$, $\mu_2 = 2 \gamma_{\rm eff}/(\hbar \omega)$, $\sigma_1 = \pm 1$ ($+$ corresponds to attractive and $-$ to repulsive self-interactions), $\sigma_2 = \Gamma/U$ and $V(x) = x^2$  is the harmonic potential due to the approximation of the pumping spots left and right of the condensate and the repulsive reservoir. 
We note that \eqref{q} is a nonlinear eigenvalue problem with real-valued eigenvalue $\mu_1$, which we will determine numerically and analytically.

\begin{figure}
\vspace{15mm}
\begin{picture}(150,70)
\put(0,-00) {\includegraphics[,width=60mm, height=40mm]{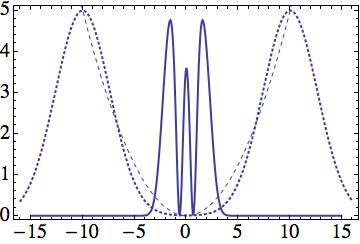} }  \put(175,100){$|\psi|^2$} \put(87,115){\textcolor{black}{x}} \put(175,85){$V$}  \put(175,72){$P$} 
\end{picture}
\caption{ 
Effective harmonic confining potential (dashed line) with $k=0.05$  vs. the pump profile $P$ (dotted line) \eqref{spots} for $A=5$, $x_2=10 \mu m$ and $x_1=-10 \mu m$. A harmonic oscillator mode forms within the trapping (continuous line). } \label{potentials}
\end{figure}
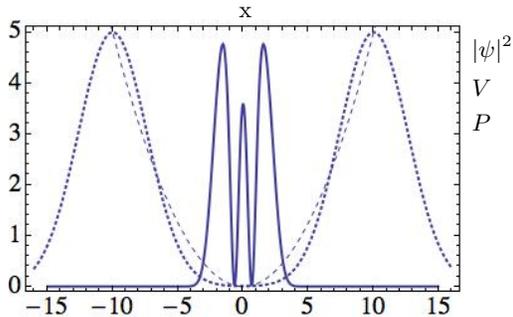
Let us identify the Carusotto-Wouters model as approximated in \cite{W, Pin8,Pin9} with the Keeling-Berloff model \cite{jona} by identifying the effective potential terms by
\beq\label{poti}
g_R N+ g_P P + \alpha_1 |\psi|^2 \leftrightarrow V(r) + U |\psi|^2
\eeq
with $V=k x^2$ and the effective gain and loss terms
\beq
 \left(\gamma_R  N - \gamma_C  \right)/2 \leftrightarrow  ( \gamma_{\rm eff} -  \Gamma |\psi|^2)
\eeq
with $N=\frac{P}{\gamma_R} (1- \frac{\beta}{\gamma_R}   |\psi|^2)$. To approximately match the potential terms, we neglect the density dependent part of $N$ (as it does not significantly affect the observed patterns) and we use a harmonic function that passes the maxima of the two Gaussian spots of \eqref{spots}. In Fig. \ref{potentials} we show the correspondence between the pump spots and the harmonic approximation and remark that $k \sim A$. To fit parameters in \eqref{poti} we note that the density dependent potential term can simply be matched $\alpha_1 = U$.

To identify the pump terms we set $\overline P - \gamma_C = 2 \gamma_{\rm eff}$, where $\overline P = 1/(|x_1-x_2|) \int^{x_2}_{x_1} P dx = 0.33 \cdot A$ is the average value of gain between both peaks assumed to be at $x_1 =-10$, $x_2 =10$ and $2 \Gamma = \beta/ \gamma_R$. Consequently the density dependent pump parameter becomes $\sigma_2 = \Gamma/U = 2.5$ while the linear parameter is $\mu_2 = (0.165 \cdot A - 0.278)/k$ with $k= m^* \omega^2 /2 = 0.092 \cdot A$ as in \cite{dark} fixing the potential strength $\omega$ via the choice of $A$.  This identification yields the simplified equation \eqref{q} once the corresponding rescaling has been applied. Recall that ${\rm sgn} (\sigma_1)$ depends on the sign of the scattering length between condensed polaritons.

By tuning the properties of the pump (in particular the geometry and pumping strength), and also having control over the nonlinear interaction strength $\sigma_1$ through the use of Feshbach resonances \cite{nature}, we see that the gain and loss coefficients ($\sigma_2$ and $\mu_2$) can be varied widely.  In the remainder of this work we will set the density dependent loss rate in rescaled units $\sigma_2 = 0.3$ and consider values for the pumping strength $\mu_1$ from $\mu_2 = 0$ to $\mu_2 = 1$, values that are in accordance with \cite{jona}. We remark that generally loss rates for polariton condensates depend significantly on the actual experimental setup, partly because lifetimes vary from $1$ps to $100$ps  and correspond to the quality of confinement within the microcavity \cite{Nat1, 1,Light, snoke,2}, so even a quasi-equilibrium scheme is feasible in polariton condensates \cite{Nat1}. This allows us to examine the equilibrium limit in our nonlinear system seamlessly ($\mu_2 = 0$), and explore the nature of the system when gain and loss are of similar magnitudes but less than or equal to the magnitude of the nonlinear term.  Setting $\sigma_1 = \pm 1$  means an increase of real-valued `nonlinear' behavior than if we used the initially introduced `natural' parameters, which is justified in physical terms by utilizing Feshbach resonances \cite{Pin11, dark,2}.

\section{The polariton harmonic oscillator}

\begin{figure}

\begin{picture}(150,70)
\put(0,-40) {\includegraphics[,width=60mm, height=40mm]{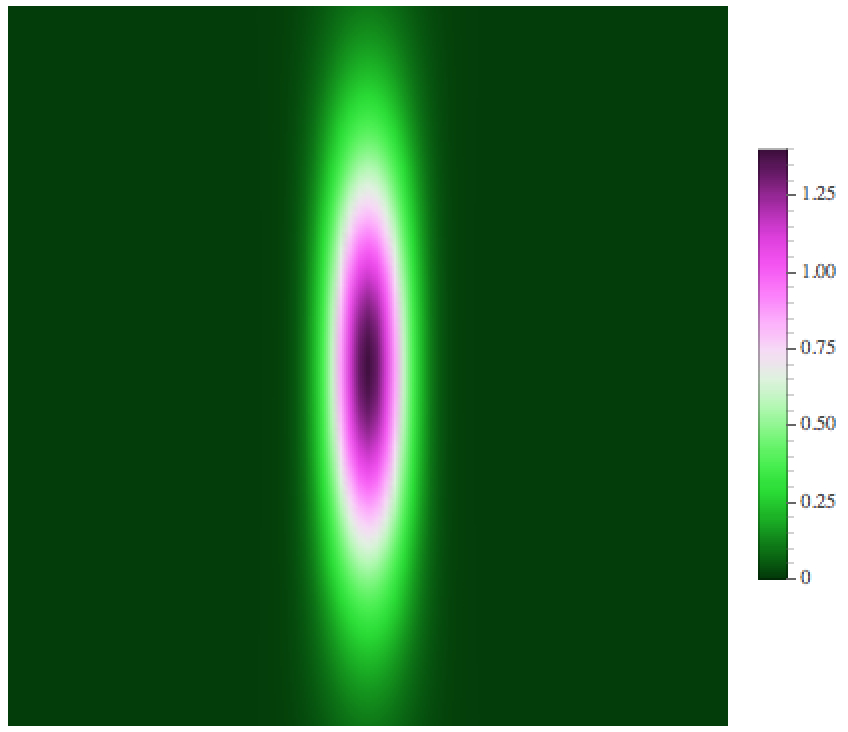} } \put(110,-25){ \textcolor{white}{(a)}}
\end{picture}
\begin{picture}(150,70)
\put(0,-90) {\includegraphics[,width=60mm, height=40mm]{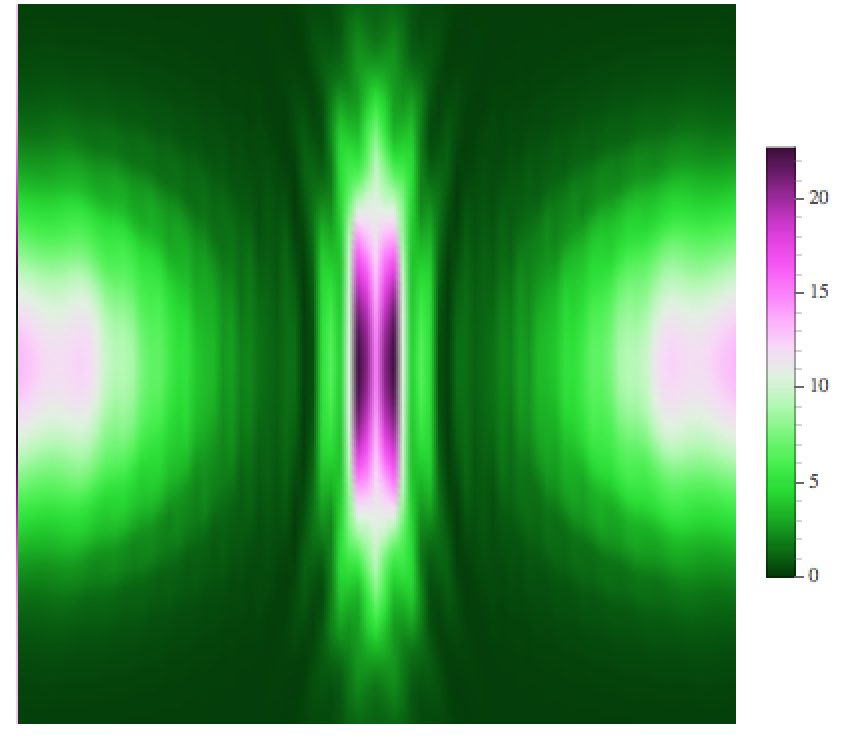} } \put(110,-75){ \textcolor{white}{(b)}}
\end{picture}
\begin{picture}(150,70)
\put(0,-140) {\includegraphics[,width=60mm, height=40mm]{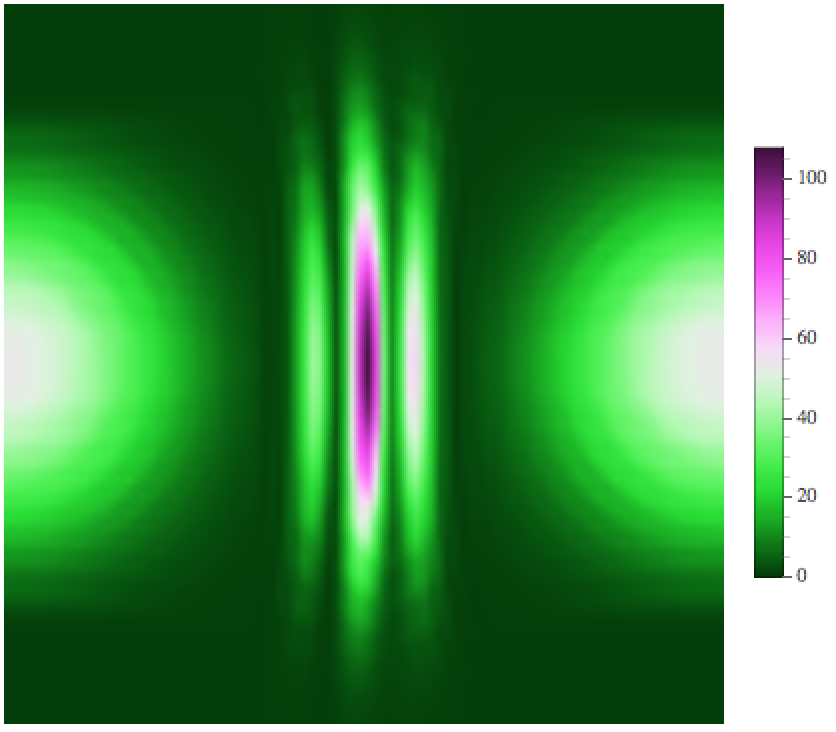} } \put(110,-125){ \textcolor{white}{(c)}}
\end{picture}
\begin{picture}(150,70)
\end{picture}
\begin{picture}(150,70)
\end{picture}

\caption{
Density profile $|\psi (x,y)|^2$ of the harmonic oscillator ground state in a $2$D setting for an area of $10 \times 10 \mu m$ is shown in (a). In (b) $|\psi(x,y)|^2$ of the first excited harmonic oscillator state in a $2$D setting for an area of $10 \times 10 \mu m$ and in (c) the density profile of the second excited harmonic oscillator state in a $2$D setting for an area of $10 \times 10 \mu m$.\cite{parameters}}\label{owl}
\end{figure}
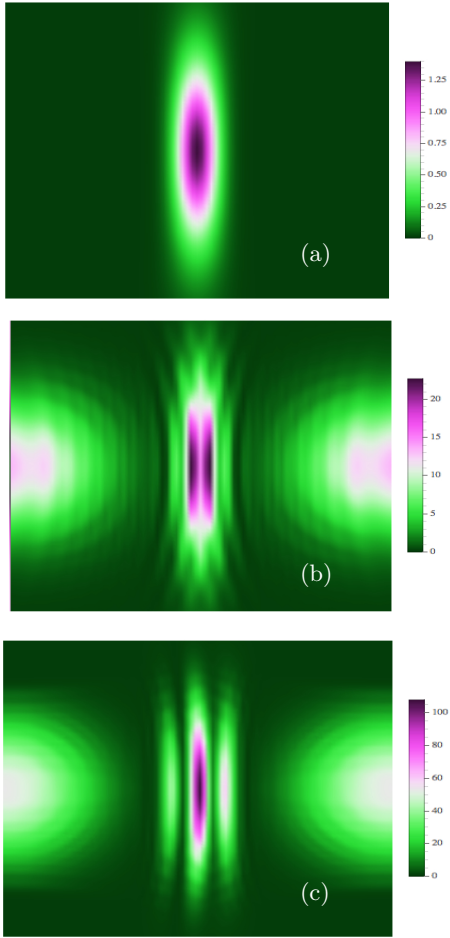

First we present results for simulations of the unreduced state equation Eq. \ref{nat}. The pump geometry is assumed to be \eqref{spots}. In  Fig. \ref{owl} (a) we present the density profile $|\psi|^2$ showing the harmonic oscillator {\it ground state} in-between two pump spots for $A=8.5$ roughly approximated by a gaussian. Fig. \ref{owl} (b) shows the {\it first} excited harmonic oscillator-type state for $A=11.5$ obeying two distinct peaks of condensate density. Finally Fig. \ref{owl} (c) shows the {\it second} excited harmonic oscillator-type state for $A=35$ obeying three distinct peaks of condensate density. Increasing the amplitude allows us to access higher order excited states.

Although we mainly aim to explain the patterns observed in \cite{Tosi}, we note that the polariton harmonic oscillator can itself be implemented within a long $1$D nanowire similar to \cite{Kamm, Pin6} i.e. a microcavity with very strong transverse confinement. Additionally an external in-plane trap can be imposed upon the effectively $1$D condensate as it has been experimentally shown for a sample consisting of three sets of four GaAs/AlAs quantum wells embedded in a GaAs/AlGaAs microcavity in \cite{ryan,ryan2}. On the other hand for the pump laser beams  there is in principle no restriction to the size of the area they are illuminating \cite{private} and thus they can be assumed to be extended over a wide range, including the area were the condensate forms within the harmonic trap. So an almost homogeneous pump distribution can be provided on the $1$D geometry on top of a repulsive and approximately harmonic trap as described by the simplified model \eqref{q}.

\section{Perturbation theory}

We now derive analytical expressions approximately solving \eqref{q}. Considering $\eqref{q}$ in the parameter limit $\sigma_1,\sigma_2, \mu_2 \to 0$ we get the well-known quantum harmonic oscillator equation
\beq\label{b}
\left( \partial_{xx} - x^2 \right) \phi_n = -E_n \phi_n,
\eeq
which has Hermite-Gauss polynomials as eigensolution
\beq
\phi_n(x)=c_n e^{-x^2/2} H_n(x)
\eeq
with $c_n = (2^n n! \sqrt{\pi} )^{-1/2}$, where 
\beq
H_n(x) = (-1)^n e^{x^2} \frac{d^n}{dx^n} e^{-x^2}
\eeq
with corresponding eigenvalues $E_n= 1+2n$ with $n \in \{0,1,2,3, \ldots \}$. These Hermite-Gauss polynomials are an orthonormal basis (ONB) of a Hilbert space.
On the other hand any wave function in a Hilbert space has an expansion in terms of Hermite polynomials of the form 
\beq\label{serious}
\psi(x) = \sum^\infty_{n=0} B_n \phi_n (x),
\eeq
with complex coefficients and real-valued functions $\phi_n (x)$ providing the ONB. By inserting this expansion in $\eqref{q}$ and  using \eqref{b} we get
\begin{multline}\label{qqq}
 \sum^\infty_{n=0} \left(  -E_n + (\sigma_1+i \sigma_2) \sum^\infty_{k=0} B_k \phi_k \sum^\infty_{l=0} B^*_l \phi_l  \right)   B_n \phi_n= \\Ê= (-\mu_1 + i \mu_2) \sum^\infty_{n=0} B_n \phi_n.
\end{multline}
By multiplying from the l.h.s. with $\phi_m$ and integrating over the whole space we obtain the formula
\begin{multline}\label{hu}
(\sigma_1+i \sigma_2) \sum^\infty_{k,l,n=0}  B_k B^*_l   B_n V_{m,k,l,n} = \\ = (E_m -\mu_1 + i\mu_2)   B_m,
\end{multline}
where we have introduced the abbreviation $V_{m,k,l,n} =  \int \phi_m \phi_k \phi_l \phi_n dx$.

We assume that for the $j$th nonlinear mode the largest coefficient in our expansion \eqref{serious} is $B_j$.  We take $B_{n<j} = 0$ and assume that only basis wavefunctions of the same symmetry of the nonlinear mode contribute in the expansion, thus we take $B_{j+2k+1} = 0$ (where  $k \in \mathbb N$).  Therefore at leading order \eqref{hu} becomes:
\begin{equation}\label{lead}
(\sigma_1+i \sigma_2) |\left|B_j\right|^2 V_{j,j,j,j} = (E_j -\mu_1 + i\mu_2).
\end{equation}
Solving simultaneously the real and imaginary parts of \eqref{lead} gives us an expression for the eigenvalue $\mu_1$,
\beq\label{mu1}
\mu_1 = E_j - \frac{\sigma_1\mu_2}{\sigma_2},
\eeq
and for $|B_j|^2$:
\beq\label{bj}
\left|B_j\right|^2 = \frac{\mu_2}{\sigma_2 V_{j,j,j,j}}.
\eeq
In our comparison with the numerical results in the next section we will use only these leading order results as they are relatively simple and easy to calculate, and immediately give us an indication of how ``nonlinear" the rescaled wavefunction is.  The eigenvalue \eqref{mu1} depends linearly on the nonlinear coefficient, but the amplitude of the rescaled wavefunction, given by the square root of \eqref{bj}, depends just on the gain and loss parameters, in a way which is in accordance with the results due to a non-conservative Lagrangian formalism presented in \cite{Pin9, Pin11}, when back scaling of the perturbation series to the wave function satisfying \eqref{n1} is made.  Thus if we see that our numerical solution agrees well with this leading order term then we can conclude that the nonlinearity plays only a minor role for the rescaled condensate wave function in this particular parameter regime.

We see that at leading order we have no information about the phase of the wavefunction.  To proceed further and find simple expressions for the higher order contributions we assume that $B_j$ is purely real and that $B_j >> B_{j+2k}$.  This allows us to find the following expression for the higher-order complex coefficients:
\beq\label{two}
B_{j+2k} = \left(\frac{\sigma_1+i \sigma_2}{E_{j+2k}-\mu_1+i\mu_2}\right)B_j^3 V_{2k,j,j,j}.
\eeq
Note that despite considering a non-equilibrium system the condensate wave function satisfies the normalization condition $\int^\infty_{- \infty} \left| \psi \right|^2 dx = \sum_k \left| B_{k} \right|^2 = c < \infty$
and $c$ corresponds to the number of condensate particles similarly to the case of fixed particle numbers \cite{Stri}.

\emph{Thomas-Fermi type regime.--} Starting from \eqref{q} we consider the regime where the kinetic energy becomes negligible compared to the external potential, nonlinearities and linear parameters, i.e.
\begin{equation}\label{q2}
\bigg(- x^2 + (\sigma_1+i \sigma_2) |\psi|^2 -  i \mu_2 \bigg) \psi \simeq   -\mu_1 \psi
\end{equation}
which to the leading order has the formal solution
\beq\label{q3}
\psi^{\rm TF} (x) = \pm \sqrt{ \frac{-\mu_1 + i \mu_2 +x^2}{\sigma_1 + i \sigma_2}}.
\eeq
Expression \eqref{q3} is however not integrable for arbitrary parameters. Besides we refer to \cite{Corr, Corr2} for a rigorous discussion of the Thomas-Fermi approximation in conservative GP theory. In contrast to the previous perturbation theory, where the coefficients \eqref{bj}  and thus the condensate wave functions dependent solely on the parameters $\mu_2$ and $\sigma_2$, the TF regime significantly depends on the nonlinearity $\sigma_1$ and the energy $\mu_1$ as formula $\eqref{q3}$ shows.


\section{Comparison between Numerical and Analytical Results}

We now proceed to test our leading order perturbation theory by comparing it with numerically calculated results.  We use a modified squared-operator method \cite{Jianke} to find numerical solutions $\phi(x)$ to the equation \eqref{q}.

In Figure \ref{sol1} we show the first three nonlinear modes for both repulsive (Fig. \ref{sol1}(a-c)) and attractive (Fig. \ref{sol1}(d-e)) interparticle interactions for the case $\mu_2 = 0.2$.  The numerical solutions are given by the thick lines, with real and imaginary parts corresponding to the dashed and dash-dotted lines respectively, while the density is given by the solid line.  We see that, as in the case of the single-component atomic BEC \cite{Kivs}, the $j$th mode has $j-1$ nodes of zero density where a $\pi$ phase change occurs.  However, unlike the atomic BEC case, the polariton condensate has a spatially dependent phase away from the phase singularity (evident through the spatially varying ratio of real to imaginary part in the wavefunction).  This variable phase indicates that particle flow must occur in the condensate, with particles moving from regions where gain dominates, to regions where loss dominates.  

We can compare the numerical results directly with the analytical expansion \eqref{serious} assuming only the leading order term is present, as given by Eqs. \eqref{mu1} and \eqref{bj}.  The analytical results are given by the thin solid lines in Fig. \ref{sol1}.  More precisely we can only compare the analytically predicted density with the numerical result, however we include also the real part and imaginary part assuming that the imaginary part is zero.  At this leading order the nonlinearity plays no role in the form of the analytical wavefunction, so it is identical for the attractive and repulsive cases.  Nevertheless, we see that it agrees remarkably well with the numerical solutions.  We see that in the repulsive case the true solution lies outside the analytical result, while the inverse is true for the attractive case, as expected.  Interestingly we see that the agreement improves as we go to higher modes, suggesting the ground state is ``more nonlinear" in the sense that it deviates more strongly from the associated linear wavefunction.

In Figure \ref{sol2} we compare the analytical and numerical results at higher gain, given by $\mu_2 = 0.5$.  We see that the pattern observed for $\mu_2 = 0.2$ becomes more evident, with the ground state showing the strongest deviation from the linear prediction.  The higher-order modes continue to show remarkably good agreement.





\begin{widetext}

\begin{figure}[ht]
\vspace{25mm}

\begin{tabular}{ccc}
\vspace{20mm}
\begin{picture}(150,50)
\put(0,0) {\includegraphics[scale=0.3]{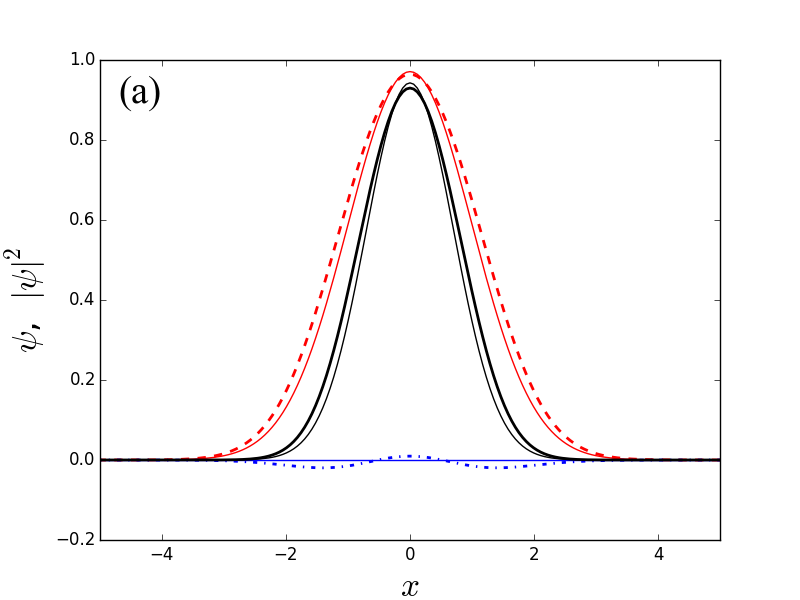} } 
\end{picture} \hspace{2mm} &
\begin{picture}(150,50)
\put(0,0) {\includegraphics[scale=0.3]{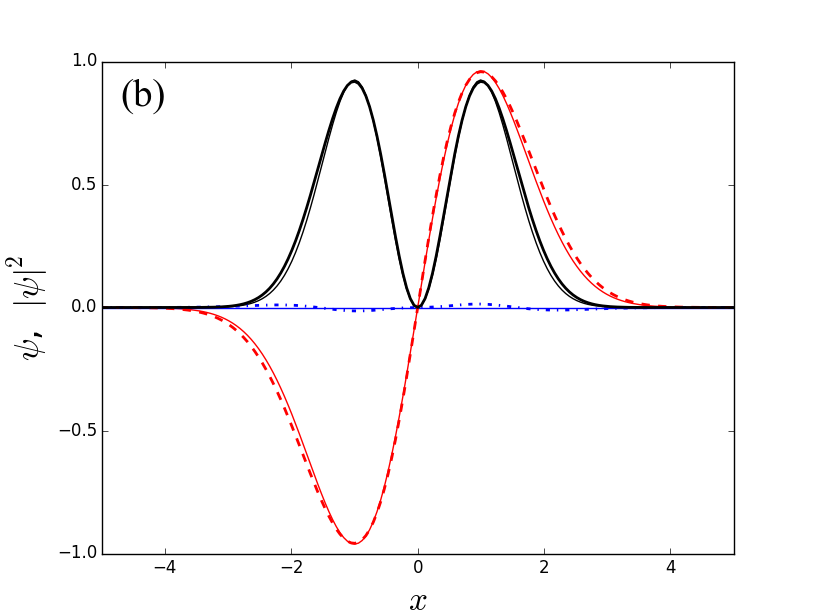} } 
\end{picture}&
\begin{picture}(150,50)
\put(10,0) {\includegraphics[scale=0.3]{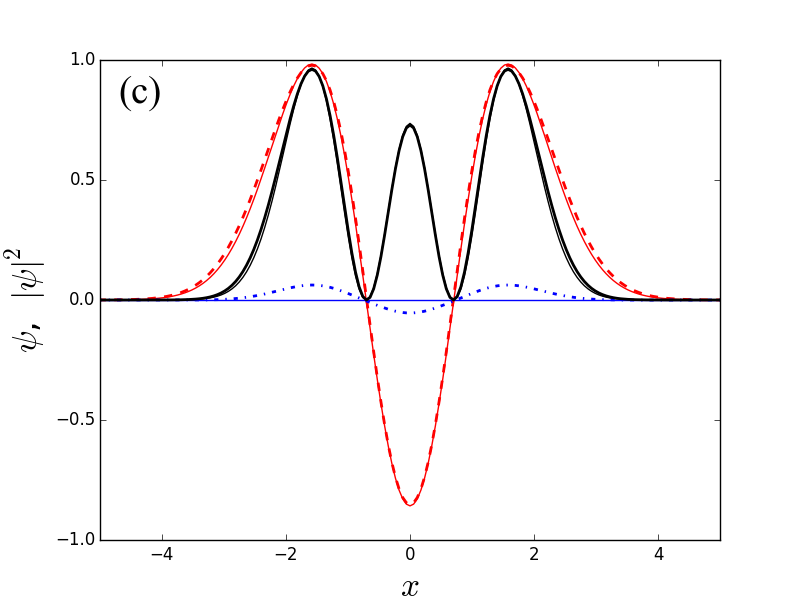} } 
\end{picture}
\end{tabular}

\vspace{0mm}

\begin{tabular}{ccc}
\vspace{0mm}
\begin{picture}(150,80)
\put(0,0) {\includegraphics[scale=0.3]{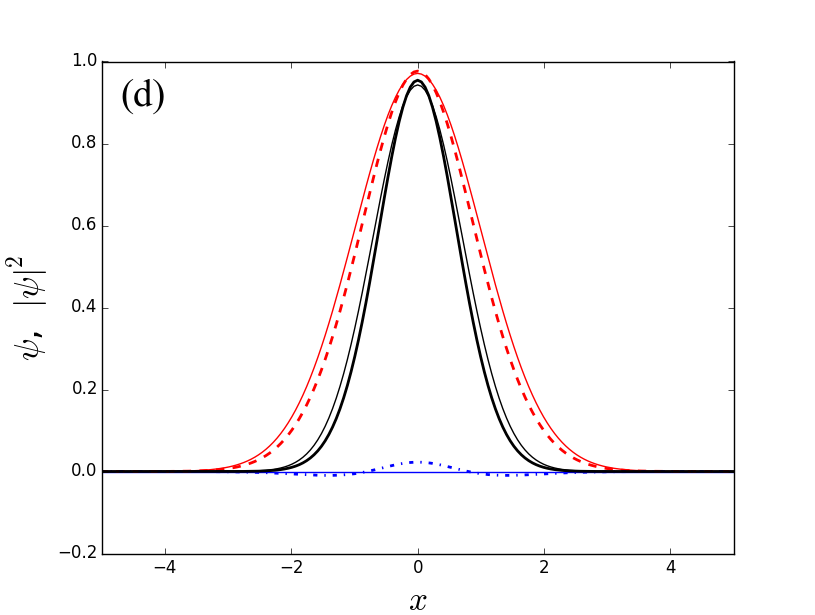} } 
\end{picture} \hspace{2mm} &
\begin{picture}(150,80)
\put(0,0) {\includegraphics[scale=0.3]{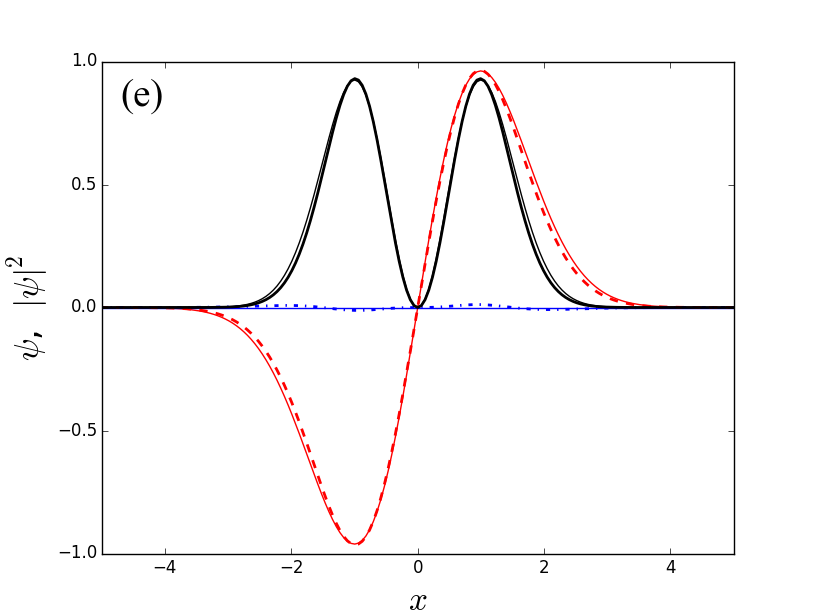} } 
\end{picture}&
\begin{picture}(150,80)
\put(10,0) {\includegraphics[scale=0.3]{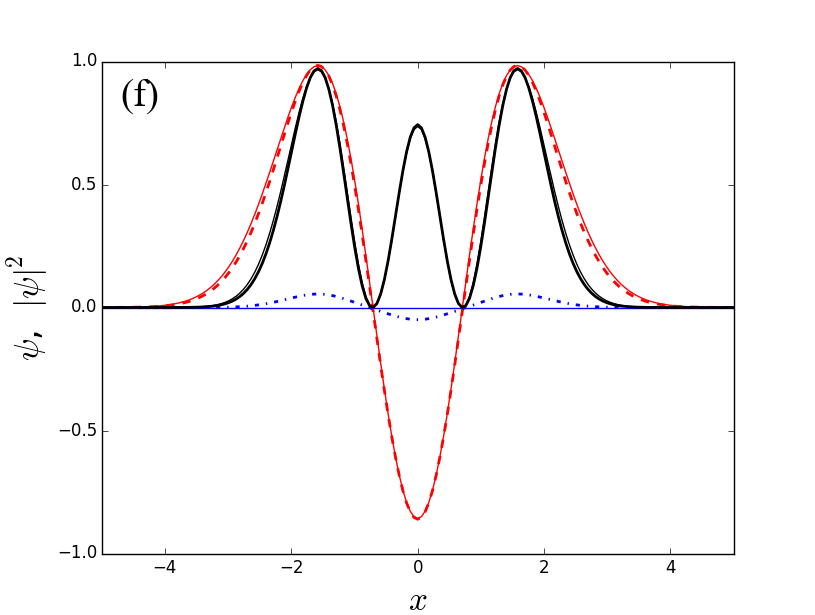} } 
\end{picture}
\end{tabular}

\caption{Comparison between numerical (thick lines) and analytical predictions (thin lines) for repulsive interactions (a-c) and attractive interactions (d-e) at gain strength $\mu_2 = 0.2$.  (a,d) Ground state; (b,e) First excited state; (c,f) Second excited state.  Numerical results follow format: density -- thick solid lines; real part -- thick dashed lines; imaginary part -- thick dash-dotted lines.  We see increasingly good agreement between the numerical and analytical predictions as we go to higher modes.}
\label{sol1}
\end{figure}

\begin{figure}[ht]
\vspace{5mm}

\begin{tabular}{ccc}
\vspace{20mm}
\begin{picture}(150,90)
\put(0,0) {\includegraphics[scale=0.3]{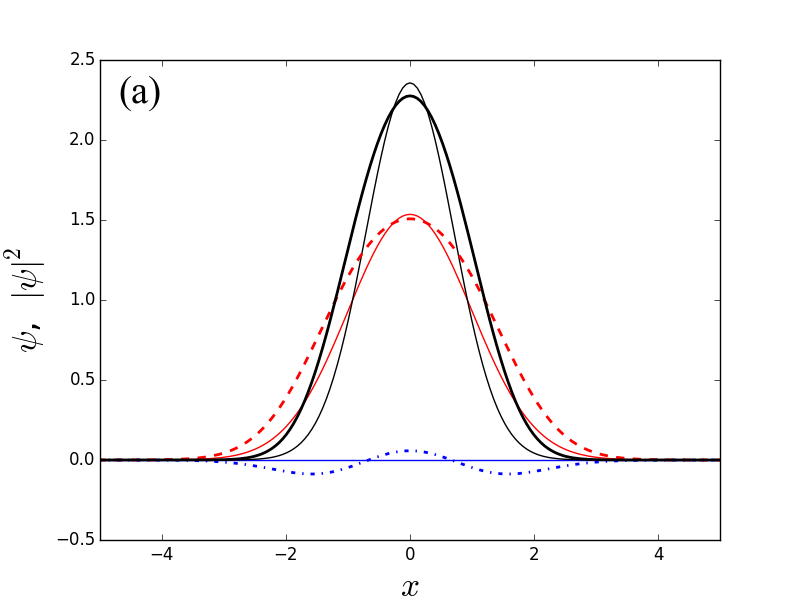} } 
\end{picture} \hspace{2mm} &
\begin{picture}(150,90)
\put(0,0) {\includegraphics[scale=0.3]{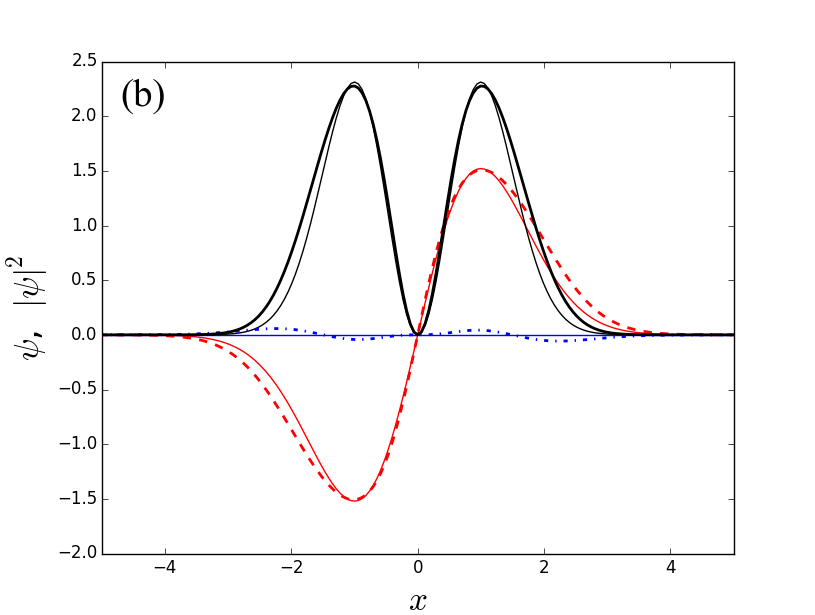} } 
\end{picture}&
\begin{picture}(150,90)
\put(10,0) {\includegraphics[scale=0.3]{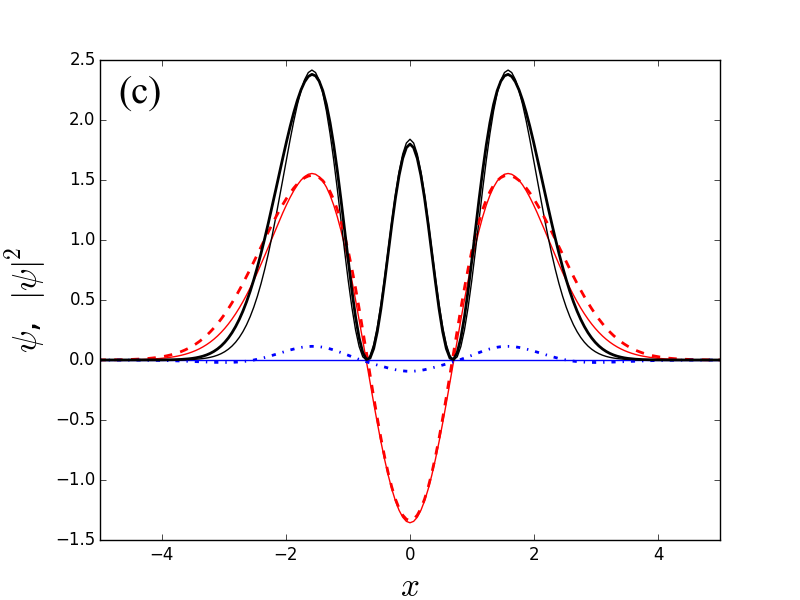} } 
\end{picture}
\end{tabular}

\vspace{20mm}

\begin{tabular}{ccc}
\vspace{0mm}
\begin{picture}(150,-30)
\put(0,0) {\includegraphics[scale=0.3]{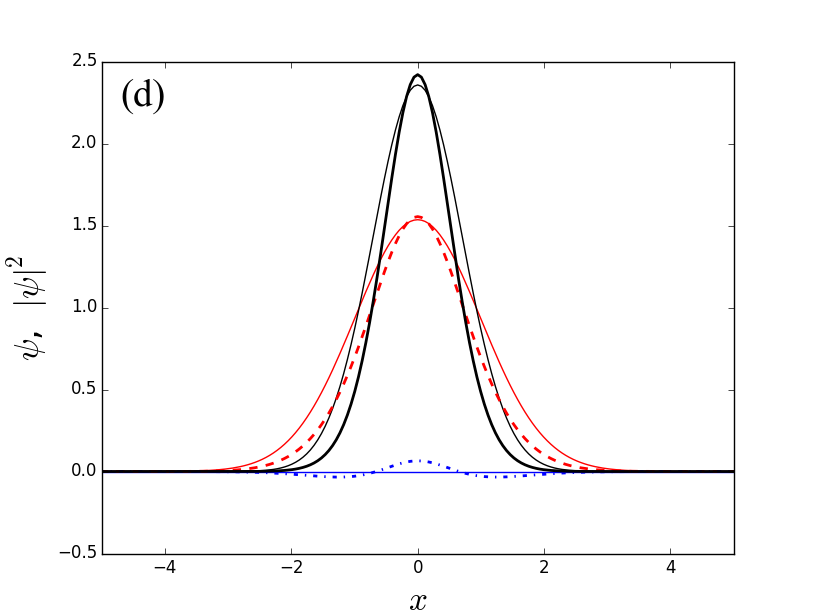} } 
\end{picture} \hspace{2mm} &
\begin{picture}(150,-30)
\put(0,0) {\includegraphics[scale=0.3]{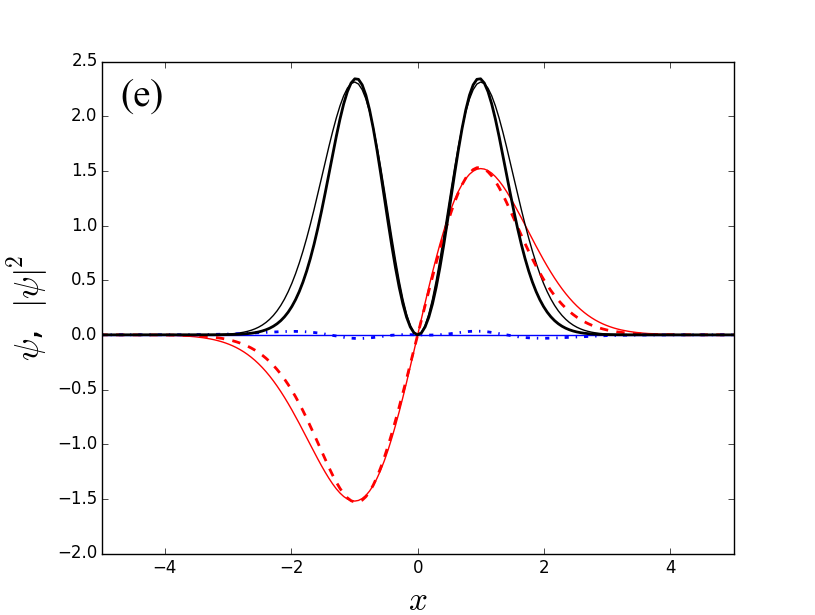} } 
\end{picture}&
\begin{picture}(150,-30)
\put(10,0) {\includegraphics[scale=0.3]{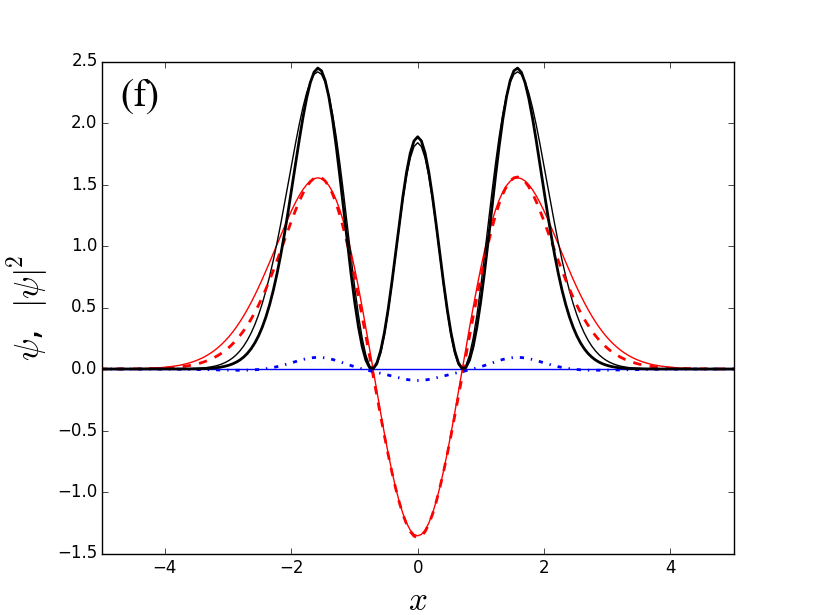} } 
\end{picture}
\end{tabular}
\caption{Comparison between numerical (thick lines) and analytical predictions (thin lines) for repulsive interactions (a-c) and attractive interactions (d-e) at gain strength $\mu_2 = 0.5$.  (a,d) Ground state; (b,e) First excited state; (c,f) Second excited state.  Format of the lines is the same as in Fig. \ref{sol1}.  We see the ground state shows significant discrepancies with the analytical predictions, however for the higher modes the agreement is still very good.}
\label{sol2}
\end{figure}

\end{widetext}

\begin{figure}
\epsfig{file=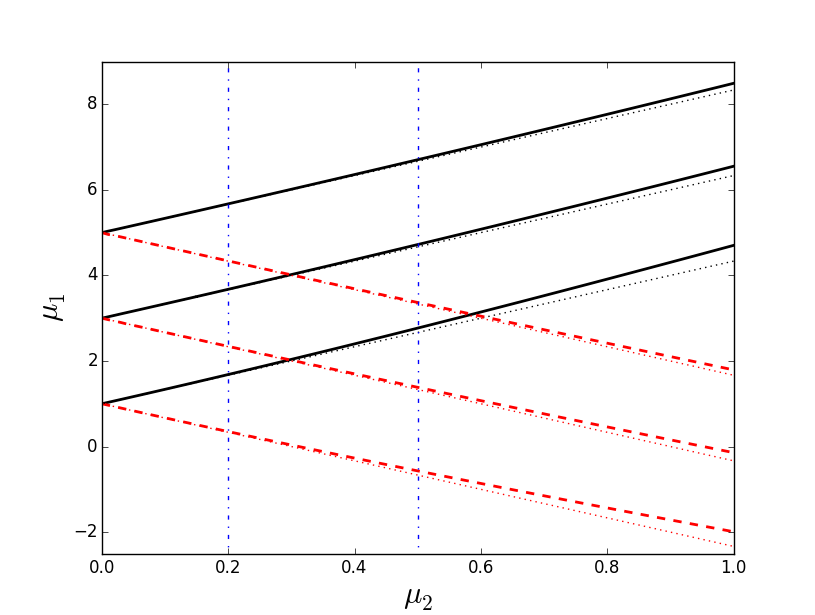,width=\columnwidth}
\caption{Family of eigenvalues, i.e. first, second and third order modes as function of pumping strength $\mu_2$.  Solid lines represent  the numerically calculated eigenvalues for the repulsive case and correspondingly the dotted lines are the analytical prediction. For attractive condensates the dashed lines show numerically generated predictions and dotted lines are the analytical predictions.}\label{eigen}
\end{figure}

The validity of the leading-order analytical prediction can be seen at a glance by comparing the predicted eigenvalues \eqref{mu1} with those found numerically.  We see in Figure \ref{eigen} the dependence of the eigenvalues on gain parameter $\mu_2$ for the first three modes (pairs of lines are arranged upper to lower and correspond to second, first zeroth modes respectively), for both repulsive (solid) and attractive (dashed) interactions.  The analytical predictions are given by the associated thin lines.  We see increasingly good agreement with mode order, with the second mode showing close agreement even up to $\mu_2 = 1$.  The ground state on the other hand is showing a discernible discrepancy by $\mu_2 = 0.5$.  The vertical dash-dotted lines correspond to the values of $\mu_2$ chosen to show the stationary states in Figs. \ref{sol1} and \ref{sol2}.  Overall we see that the features of the stationary states are dictated predominantly by the gain and loss coefficients, rather than the interparticle interactions.

\section{NOR gate and a {\lowercase{n}}NOR gate} 
The pattern formation within a $1$D nanowire can have the following technological application as NOR gate. A NOR gate \cite{nor}  is defined by the logic that two or more zero inputs yield an output $1$ in a boolean logic system. Here we can identify each pump spot \eqref{spots} with $0$ or $1$ by defining that, if the laser beam is turned on, $A \neq 0$, the input corresponds to $0$ and if not it corresponds to $1$. The output will be the quantum harmonic oscillator patterns observed, iff both pump spots are equal $0$. So we can identify all the harmonic oscillator patterns with an output $1$, and thus we have established a simple feasible NOR gate by means of topologically stable excitations carried by a non-equilibrium polariton condensate within a semiconductor microcavity which stand in the tradition of optical computing \cite{Pin10}.


This NOR gate  can be generalized in the sense that we do not solely distinguish between "on" corresponding to $0$ and "off" associated with $1$ of both pump spots, but in addition include the experimental and numerical observation that different pump strengths lead to different numbers of density lobes. Therefore we can enumerate the logical $0$s leading to different lobes states, i.e. ground state, first excited state, second excited state and so on. That is, if the pump spots have a certain amplitude $A$ it leads to patterns of say $B(A): \mathbb R \to \mathbb N$ lobes.  However when only one pump spot is turned on, corresponding to a logical input, $0$ and $1$, or boths pumps are off, corresponding to $0$ and $0$ input, no lobes patterns form \cite{Kamm2}.

\section{Conclusions}

Motivated by a recent experiment we considered a mean field model for polariton condensates to reproduce the observed patterns. Starting from this we could identify the general model with that of a simpler nonlinear non-equilibrium quantum harmonic oscillator.
For this effective theory we have presented a perturbative approach and so generalized the quantum harmonic oscillator theory to include the non-conservative character of the non-equilibrium polariton condensate. We introduced the family of excited states starting from the gaussian ground state to the $n$ node non-equilibrium quantum harmonic oscillator states.
The analytical eigensolutions were compared to corresponding numerical simulations and showed excellent agreement. Due to the recent experimental accessibility of Feshbach resonances in polariton systems  we further predicted  the scenario of attractive self-interactions and outlined their role in nonlinear pattern formation.  Particularly the modification of the density formation was explicitly illustrated.
In the course of our analysis we showed that the energy eigenvalues depend on the pumping strength $\mu_2$ linearly  to the leading order  and conclude to provide a thorough explanation for the experimentally observed patterns.   Finally we discussed the possibility for a generalized optical NOR device in polariton condensates utilizing the underlying pattern formation logic.

\section{Acknowledgements}

FP has been financially supported through his EPSRC doctoral prize fellowship at the University of Cambridge. FP likes to thank Natasha Berloff, Hamid Ohadi and Alexander Dreismann for stimulating discussions.





\begin{thebibliography}{10}
\bibitem{parameters} The time evolution has been induced via a $4$th order Runge-Kutta method and a $4$ order finite differences scheme. We used a computational window of $226$ space units in $x$ and along the transverse directions $y$. Typical space steps are about $\Delta x=0.1$ and the time steps $\Delta t \simeq 1.5\times 10^{-4}$.
\end{thebibliography}
  \end{document}